\documentclass[pdflatex,sn-mathphys-num]{sn-jnl}

\usepackage{graphicx}%
\usepackage{multirow}%
\usepackage{amsmath,amssymb,amsfonts}%
\usepackage{amsthm}%
\usepackage{mathrsfs}%
\usepackage[title]{appendix}%
\usepackage{xcolor}%
\usepackage{textcomp}%
\usepackage{manyfoot}%
\usepackage{booktabs}%
\usepackage{algorithm}%
\usepackage{algorithmicx}%
\usepackage{algpseudocode}%
\usepackage{listings}%
\usepackage{aas_macros}

\usepackage{hyperref}

\theoremstyle{thmstyleone}%
\theoremstyle{thmstyletwo}%

\theoremstyle{thmstylethree}%

\begin{document}

\title[Article Title]{Dark matter alters dwarf galaxy sizes and morphologies}

\author*[1,2,3]{\fnm{Akaxia} \sur{Cruz}}\email{akaxia@princeton.edu}

\author[2]{\fnm{Robel} \sur{Geda}}\email{robel@princeton.edu}

\author[4]{\fnm{Alyson} \sur{Brooks}}\email{abrooks@physics.rutgers.edu}

\author[1,3]{\fnm{Mariangela} \sur{Lisanti}}\email{mlisanti@princeton.edu}

\author[5]{\fnm{Annika H. G.} \sur{Peter}}\email{peter.33@osu.edu}

\author[6]{\fnm{Thomas R.} \sur{Quinn}}\email{trq@astro.washington.edu}

\author[7]{\fnm{Ben} \sur{Keller}}\email{bkeller1@memphis.edu}


\affil*[1]{\orgdiv{Department of Physics}, \orgname{Princeton University}, \orgaddress{\street{Jadwin Hall, Washington Rd}, \city{Princeton}, \postcode{08544}, \state{New Jersey}, \country{USA}}}

\affil[2]{\orgdiv{Department of Astrophysical Sciences}, \orgname{Princeton University}, \\\orgaddress{\street{Peyton Hall, 4 Ivy Lane}, \city{Princeton}, \postcode{08544}, \state{New Jersey}, \country{USA}}}

\affil[3]{\orgdiv{Center for Computational Astrophysics}, \orgname{Flatiron Institute}, \\\orgaddress{\street{5th Ave}, \city{New York}, \postcode{10010}, \state{New York}, \country{USA}}}

\affil[4]{\orgdiv{Department of Physics and Astronomy}, \orgname{Rutgers University}, \orgaddress{\street{Frelinghuysen Rd}, \city{Piscataway}, \postcode{08854}, \state{New Jersey}, \country{USA}}}

\affil[5]{\orgdiv{Department of Physics and Astronomy}, \orgname{The Ohio State University}, \\\orgaddress{\street{191 W. Woodruff Ave}, \city{Columbus}, \postcode{43210}, \state{Ohio}, \country{USA}}}

\affil[6]{\orgdiv{Department of Astronomy}, \orgname{University of Washington}, \orgaddress{\street{15th Ave NE}, \city{Seattle}, \postcode{98195}, \state{Washington}, \country{USA}}}

\affil[7]{\orgdiv{Department of Physics and Materials Science}, \orgname{University of Memphis}, \orgaddress{\street{216 Manning Hall}, \city{Memphis}, \postcode{38152}, \state{Tennessee}, \country{USA}}}



\abstract{Dwarf galaxies are the most dark-matter-dominated systems in the Universe~\cite{Mateo1998, Moster2013}, serving as unique laboratories for exploring the fundamental nature of the dark sector. Although the observed diversity in dwarf galaxy sizes and morphologies~\cite{Lelli2016} suggests an interplay between dark matter and baryonic physics~\cite{Lelli2022}, this relationship has remained largely unquantified in existing models. Here, we utilize a set of novel cosmological simulations with baryons to demonstrate how dark matter physics directly drives the structural evolution of these galaxies. By comparing the same systems across cold dark matter~\cite{Cruz2026} and varied self-interacting dark matter models, we find that the evolution of the central density of dark matter significantly impacts a galaxy's half-light radius, but the signal is dependent on morphology. This causal link establishes galaxy size as a direct signature of the underlying dark matter microphysics. Our findings provide a predictive framework for decoding dark matter particle properties from galactic shapes, offering a roadmap for upcoming wide-field surveys, such as the Legacy Survey of Space and Time~\cite{LSST2019}, to probe the dark sector through the statistical analysis of galaxy populations. }

\keywords{Cosmology, Dark Matter, Dwarf Galaxies, Galaxy Formation}

\maketitle

The physics setting a galaxy's spatial extent remains poorly understood. State-of-the-art hydrodynamical simulations are only now achieving the resolution and sample size required to isolate the mechanisms driving dwarf galaxy structure. While earlier cold dark matter~(CDM) simulations failed to reproduce the observed size--stellar mass~($M_*$) relation in dwarf galaxies or link halo spin~\cite{Dalcanton1997, Mo1998, Fall1980, White1984, Maller2002, Zavala2008} to galaxy size~\cite{Teklu2015, Zjupa2017, Jiang2019b, VanderWel2011, Carlsten2021, Jiang2019, Sales2022}, the recent Massive Marvelous Dwarf suite~\cite{Cruz2026}, which directly targets the mass regime where diversity is maximized~\cite{Oman2015}, successfully captures the scatter in stellar size at fixed mass. It also shows, for the first time, that orbital angular momentum from mergers drives dwarf disk formation through gas interactions~\cite{Geda2026}. This paper extends the Massive Dwarf suite by re-simulating a subset of its galaxies using a self-interacting dark matter~(SIDM) model~\cite{Spergel2000}, with galaxy formation physics held constant. The resulting Massive SIDM Suite Investigating Dwarf Morphology~(SIDM-SIDM) zoom-ins represent the largest, highest-resolution hydrodynamical sample of dwarf-mass simulations for the low and high SIDM cross sections explored here, offering a unique opportunity to test how dark matter physics shapes galaxy size and morphology.

Distinct from CDM, SIDM halos undergo two main evolutionary phases. First, their density profiles form low and constant-density dark matter cores.  This phase has been considered a solution to the core-cusp problem~\cite{Dave2001, Rocha2013, Fry2015} and was the focus of previous hydrodynamical studies of dwarf-mass SIDM halos~\cite{Fitts2019,Robles2017}.  During the second phase, the central density undergoes runaway collapse due to the negative heat capacity of the halo's self-gravitating system~\cite{Balberg2002}. In contrast to CDM, where the conduction rate is set by the Chandrasekhar relaxation time due to random two-body gravitational collisions~\cite{Chandrasekhar1942}, the SIDM enhanced conduction rate, which depends on the self-interaction cross section, can cause collapse within a Hubble time. The timescale to reach the core-collapse phase in dark-matter-only simulations depends on the SIDM cross section, dark matter halo mass, and concentration~\cite{Essig2019, Zeng2022}, as well as merger history~\cite{2026arXiv260602566S}.

We select a subset of 24 halos from the Massive Marvelous Dwarf suite~\cite{Cruz2026}, with $M_{\rm 200} \sim 1.4 \times (10^{10}\text{--}10^{11})~{\rm M}_{\odot}$ and $M_* \sim (5 \times 10^7)\text{--}(3 \times 10^9)~{\rm M}_{\odot}$. These 24 initial conditions are re-simulated using a velocity-dependent SIDM model~(VX-SIDM) based on a Yukawa interaction~\cite{Feng2010, Loeb2011, Tulin2013}. We additionally consider a velocity-independent model~(SI50) with a constant transfer cross section of 50~cm$^2$/g, which is high enough to trigger core collapse in the 9 zoom-in galaxies it is run on. The SI50 model is only examined up to $z = 0.3$ due to convergence limits on the central density evolution in the core-collapse regime and computational resources. Model details, including agreement with current constraints, simulation parameters, and convergence testing results are provided in the Methods Section~\ref{methods} and in the Appendix~\hyperref[sec:convergeB]{B}. For a given initial condition, halo mass and concentration remain roughly fixed. Therefore, altering the SIDM cross section determines whether the halo retains a core or enters core collapse by redshift $z = 0$.

Leveraging the Massive Marvelous and SIDM-SIDM suites, we systematically compare galactic structure across dark matter models to isolate how and when stellar distributions co-evolve with dark matter cores. By doing so, we find a correlation between the stellar half-light radius and the dark matter core radius for some galaxies, demonstrating how specific merger histories can disrupt this correspondence for others. These results lay the groundwork for using galaxy morphology as a direct probe of dark matter microphysics.

\section{Co-evolution of dark matter cores and galaxy sizes}\label{sec:coevol}

\begin{figure}[t]
\centering 
\includegraphics[width=\linewidth]{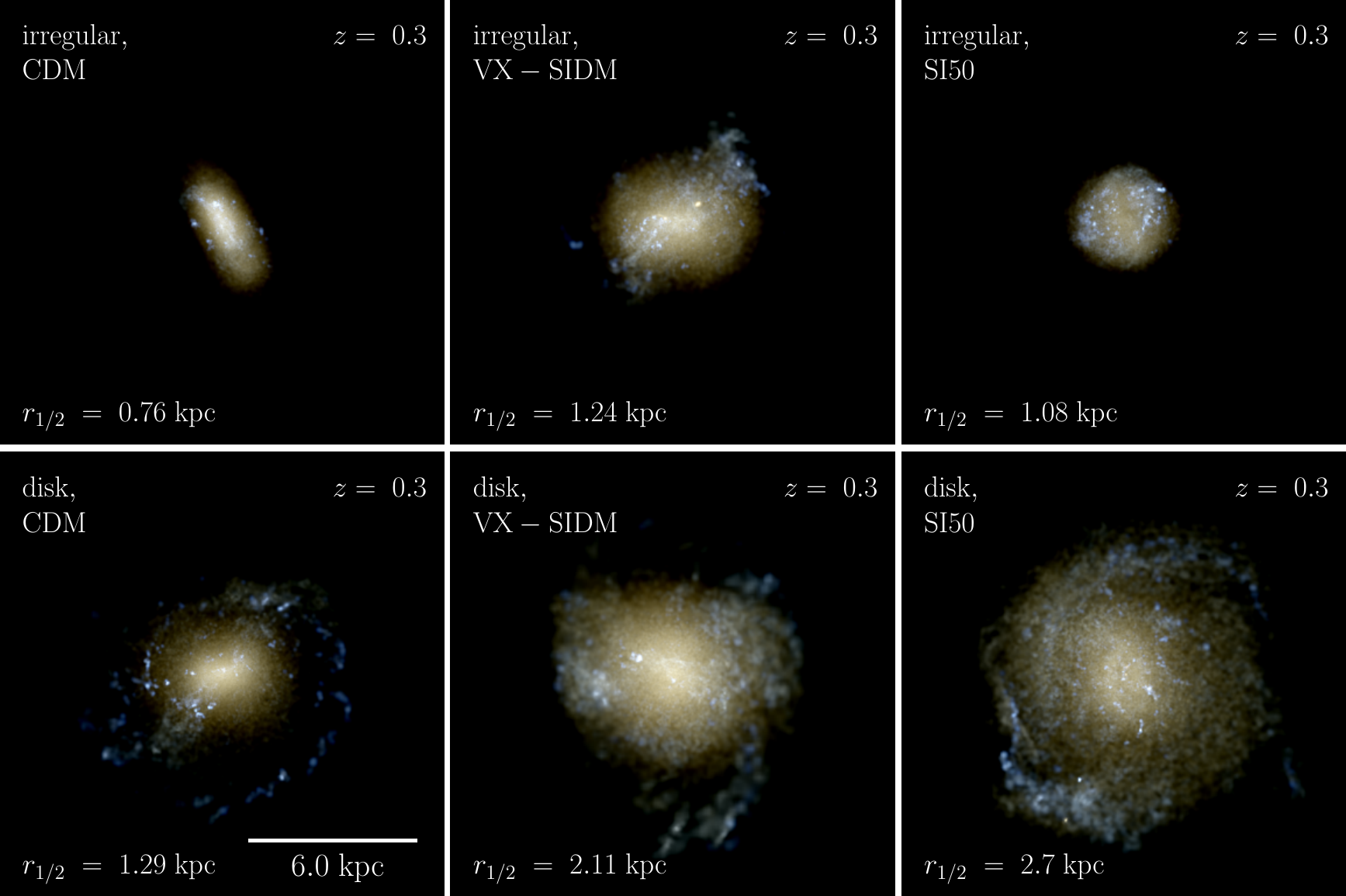}
\caption{Mock false color UVI images of two representative galaxies at redshift $z=0.3$ for the three different dark matter cosmologies: CDM~(left), VX-SIDM~(middle), and SI50~(right). The stellar size is annotated in the bottom-left corner of each panel.  {\bf Top}: A representative irregular galaxy. In CDM, it exhibits a bar-like structure. In VX-SIDM, it appears visually larger than in CDM, while in SI50, it is more compact than in VX-SIDM and more symmetric in the x-y plane than in the other cosmologies. {\bf Bottom}: A representative disk galaxy. Disks form in all three cosmologies, with CDM producing the smallest one.}
\label{fig:uvi}
\end{figure}

We first explore how dark matter physics influences the size of two representative galaxies in the suite. Figure~\ref{fig:uvi} shows mock face-on, false color, ultraviolet, visible, and infrared~(UVI) filter images of an irregular system~(top row) and a disk~(bottom row). Stellar sizes are measured using the $K$-band half-light radius ($r_{1/2}$) \cite[see][]{Cruz2026}. A galaxy is classified as having a disk morphology if its stellar population has a normalized, median specific angular momentum in the z-direction greater than 0.5~\cite{Cruz2026,Geda2026}; irregular galaxies, which include spheroidals, have a value less than 0.5. From left-to-right, the columns demonstrate how each galaxy changes across the CDM, VX-SIDM, and SI50 cosmologies at $z = 0.3$. The $r_{1/2}$ of the irregular galaxy increases by $\sim 63\%$~(42\%) relative to CDM for the VX-SIDM~(SI50) cosmology. The disk galaxy in the bottom row shows similar variations with cosmology, with its size increasing by $\sim 64\%$~(109\%) for VX-SIDM~(SI50). A notable difference is that the disk galaxy has the largest $r_{1/2}$ in the SI50 cosmology, while the irregular galaxy has the largest $r_{1/2}$ for VX-SIDM. 

\begin{figure}[t]
\centering
\includegraphics[width=1\linewidth]{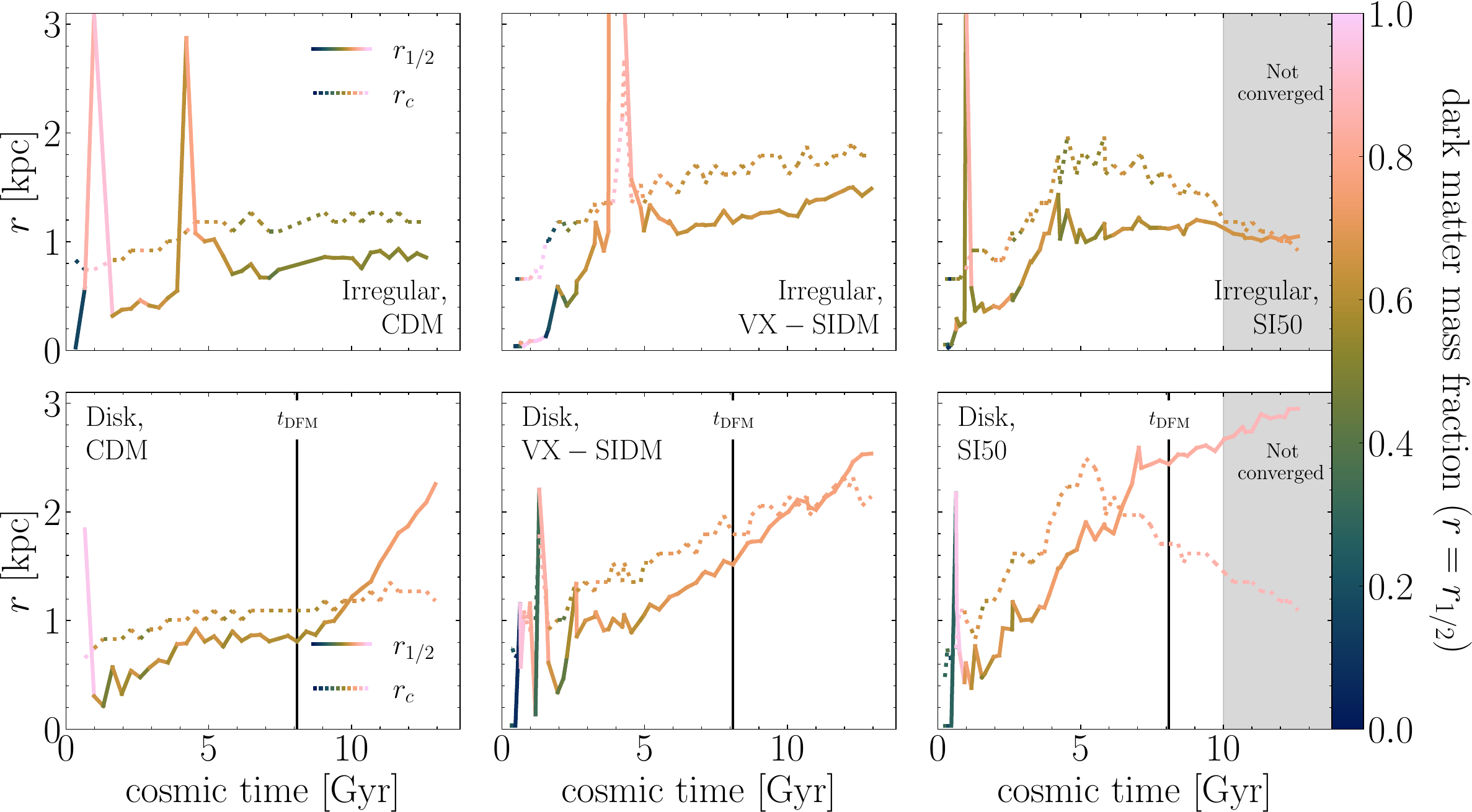}
\caption{The time evolution of dark matter core size ($r_c$, dotted lines) and galaxy half-light radius ($r_{1/2}$, solid lines), with tracks colored by the dark matter mass fraction within $r_{1/2}$. Results are shown for a representative irregular~(top row) and disk~(bottom row) galaxy in three different cosmologies: CDM~(left), VX-SIDM~(middle), and SI50~(right). In the SI50 model, the gray region corresponds to times where the central density evolution is not converged for the time-stepping scheme used. In both cases, dark matter either dominates or exists in roughly equal proportion to the baryons within $r_{1/2}$, except at early times near spikes (mergers) where the galaxy and dark matter can decouple. {\bf Top}: For the irregular galaxy in all three cosmologies, $r_c$ and $r_{1/2}$ co-evolve but follow distinct evolutionary trajectories driven by each dark matter cosmology.  {\bf Bottom:} Similarly, for the disk galaxy in all three cosmologies, $r_{1/2}$ and $r_c$ co-evolve until the disk-forming merger at time $t_{\scriptscriptstyle{\rm DFM}}$. In all cases, the minimum obtainable core size is limited by the CDM convergence radius $r_{\rm conv} = 0.45$~kpc.} \label{fig:size_evol}
\end{figure}

These examples illustrate that a galaxy's morphology (irregular versus disk) strongly influences how its low-$z$ stellar size corresponds to dark matter cosmology. Figure~\ref{fig:size_evol} shows that these differences are governed by the halo's formation history and the evolution of its dark matter core. To demonstrate this, the figure plots the core~size ($r_c$, dotted line) and stellar size ($r_{1/2}$, solid line), where $r_c$ is the radius at which the maximum resolved dark matter density ($r_{\rm conv}$, see Methods Section~\ref{methods}) falls by a factor of two. The line color tracks the dark matter mass fraction within $r_{1/2}$ over time, though this may become inaccurate during violent mergers when $r_{1/2}$ spikes and the galaxy and dark matter may decouple.

In CDM, the irregular galaxy forms a $\sim 1$~kpc dark matter core early due to bursty supernovae feedback~\cite{Pontzen2012}, and its size remains essentially constant until present day following an initially turbulent period with major mergers. The core sizes found in the Massive Marvelous Dwarfs are consistent with~\cite{Pontzen2012}, which compared an analytic model to CDM simulations and predicted $\sim 1$~kpc cores from gravitational fluctuations from stellar feedback, for systems of similar mass to those studied here. The stellar half-light radius follows a similar trajectory, reaching a roughly constant value of 0.84~kpc. Starbursts drive potential fluctuations that transfer energy to collisionless particles, expanding dark matter orbits into central cores. Because stars are also collisionless, their orbits expand in tandem with the dark matter, driving $r_{1/2}$ and $r_c$ to co-evolve. In fact,~\cite{Pontzen2012} verified that the collisionless stars in their simulations also obtained expanded orbits, but did not directly examine their evolution~\cite[see also][]{Riggs2024}. Because CDM core formation is driven by baryons, the feedback prescription may alter the evolution of core size~\cite[but see][for how this may make observationally inconsistent stellar sizes]{Cruz2026}. \\

In SIDM halos, core evolution follows a different path: the core initially expands as central density drops, then shrinks rapidly once collapse begins. Because the VX-SIDM cross section is insufficient to trigger collapse within a Hubble time, $r_c$ grows  monotonically to 1.71~kpc at $z=0$, excluding mergers. In contrast, the SI50 model has a large-enough cross section to drive collapse. The irregular galaxy's $r_c$ grows to a maximal size of 1.97~kpc at 5.83~Gyr before starting to shrink. As in CDM, $r_{1/2}$ tracks the $r_c$ evolution in both SIDM models: it grows continuously in VX-SIDM but eventually reverses course in SI50. Consequently, the low-$z$ irregular galaxy is larger in VX-SIDM than in SI50. Although feedback is also present in the SIDM galaxies, Figure~\ref{fig:size_evol} shows that SIDM thermalization plays the primary role in driving $r_{ 
1/2}$ evolution in VX-SIDM and during the early stages of SI50. At later times in SI50, however, feedback may prevent a further decrease in stellar size. In the representative irregular galaxy, the dark matter mass fraction within $r_{1/2}$ indicates that dark matter dominates or is in equal abundance to baryons across the dwarf's evolution, indicating that the co-evolution between the stars and dark matter is driven by the dark matter's gravitational potential. Using Jeans equations in the limit of a dark-matter-dominated, static, isothermal, spherical system $r_c \sim \mathcal{O}(1) ~r_{1/2,{\rm b}}$ where $r_{1/2, {\rm b}}$ is the half-mass radius of baryons. Within $r_c$, HI gas should dominate the gas mass budget and a correlation between $r_{1/2}$ and $r_{\rm HI}$, the HI radius~\cite{Cruz2026}, exists. The $r_c$ and $r_{1/2}$ co-evolution driven by dark matter is distinct from the mechanism described in~\cite{Kaplinghat2014} where instead baryons drive the dark matter size in more massive galaxies. However in the full population of irregular galaxies there are instances where the dark matter mass fraction is less than 0.5 (dark green/blue) at low-$z$, indicating that the systems have become baryon dominated. Such irregular systems (see e.g., r468, r488, r556 in the Appendix~\hyperref[sec:coevol_A]{A}) may also be subject to the mechanism of~\cite{Kaplinghat2014} where baryons determine the dark matter core size at low-$z$. 

Crucially, how stars respond to a halo's changing gravitational potential depends on the galaxy's formation history, specifically whether it forms a disk.  This is illustrated by the disk galaxy example in Figure~\ref{fig:size_evol}. The co-evolution between $r_c$ and $r_{1/2}$ proceeds similarly to that of the irregular galaxy until the disk forms. At $t_{\scriptscriptstyle{\rm DFM}}$, the galaxy undergoes a disk-forming merger~(DFM), which increases the angular momentum of the gas and initiates disk growth~\cite{Geda2026}. In particular, disk formation can occur after a gas-rich merger with high orbital angular momentum, which increases the radial extent of the central galaxy's surrounding gas. Star formation in the resulting gas disk then yields the steady growth of an angular-momentum-supported stellar population, gradually increasing $r_{1/2}$. The disk is maintained by additional low pericenter, co-aligned, gas-rich mergers or prolonged merger quiescence longer than 2~Gyr. In contrast, the irregular galaxies undergo misaligned or head-on mergers throughout their formation histories, which prevent disk formation \citep[see][for a detailed analysis of dwarf disk formation in CDM]{Geda2026}. After $t_{\scriptscriptstyle{\rm DFM}}$, $r_c$ maintains its trajectory---remaining constant in CDM, growing in VX-SIDM, and shrinking in SI50. However, $r_{1/2}$ diverges from this evolution and begins to increase. The deviation in co-evolution is obvious for the CDM and SI50 cosmology. It is less obvious in the VX-SIDM case due to the large $r_c$, but the fact that the rate of increase in $r_{1/2}$ surpasses that of $r_c$ during disk formation suggests that this process is at play.  

In short, $r_c$ and $r_{1/2}$ co-evolve if there is no merger-driven disk formation. If instead a disk forms, the $r_c$ and $r_{1/2}$ co-evolve until the high-angular-momentum merger.

\section{Core \& galaxy size relationship at low redshifts}\label{low_z_sizes}

\begin{figure}[t]
\centering
\includegraphics[width=\linewidth]{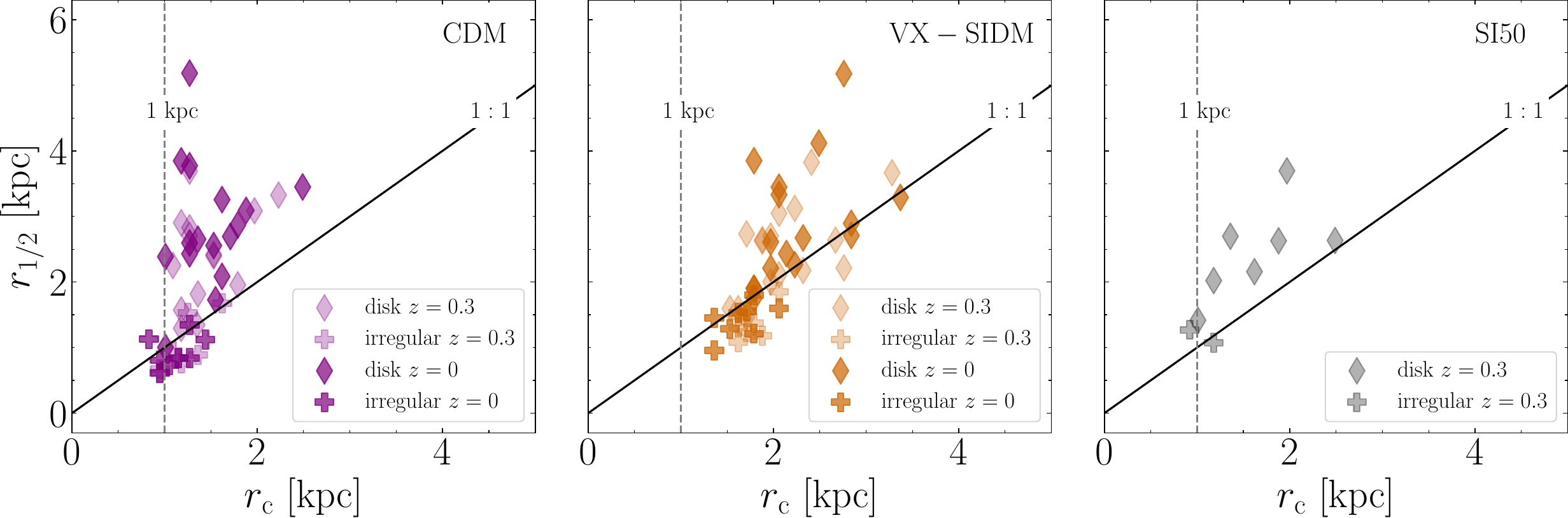}
\caption{A comparison between $r_c$ and $r_{1/2}$ for full simulated suites at $z = 0$ (opaque markers) and at $z = 0.3$ (transparent markers). In all three cosmologies, the irregular galaxies sit near the 1:1 line~(black), yet the irregular galaxies in CDM are more compact than in VX-SIDM, and similarly as compact for the core-collapsing SI50. {\bf Left:} Given the constant $r_c$ sizes through time in CDM (purple) there is a vertical relationship between $r_c$ and $r_{1/2}$ at both redshifts, emphasized with the dashed gray 1 kpc line. The larger galaxy sizes are set by disk formation, which grow from $z = 0.3$ to $z = 0$. {\bf Middle:} At both redshifts, the VX-SIDM (orange) instead has a nearly 1:1 relationship, excluding the largest disks. This is driven by the continued growth of $r_c$ as $r_{1/2}$ increases in the presence of disk formation. {\bf Right:} In SI50 (grey), shown at $z = 0.3$, the slope of the relation is in between CDM and VX-SIDM, driven by $r_c$ becoming smaller due to core collapse.} \label{fig:rc_rh}
\end{figure}

The co-evolution of $r_c$ and $r_{1/2}$ of galaxies before $t_{\scriptscriptstyle{\rm DFM}}$, and the break after, generalizes across the entire simulation suite. Figure~\ref{fig:rc_rh} shows the relationship between $r_c$ and $r_{1/2}$ at low-$z$ for all galaxies in the suite. The CDM and VX-SIDM results are shown at both $z = 0.3$~(transparent markers) and $z=0$~(opaque markers), while the SI50 ones are only shown at $z = 0.3$ because core collapse is not converged at later times in the simulations. Across all three cosmologies, irregular galaxies follow a scaling near $1:1$ between $r_c$ and $r_{1/2}$. At $z = 0.3~(0)$, the average ratio of $r_{1/2}$ to $r_c$ is 0.81 $\pm$ 0.20 (0.85 $\pm$ 0.23) in CDM, 0.79 $\pm$ 0.10 (0.86 $\pm$ 0.13) in VX-SIDM, and 1.15 $\pm$ 0.24 in SI50. Generically, the core sizes (and thus galaxy sizes) at present day are larger in VX-SIDM compared to those in CDM, driven by the larger cores formed in that model. The SI50 galaxies instead have core sizes more consistent with CDM due to core collapse and a larger ratio of $r_{1/2}$ to $r_{c}$ due to a lag in shrinking galaxy size after maximal core formation. 

In contrast, disk galaxies display distinct, cosmology-dependent behavior at low-$z$. In CDM, galaxies have $r_c\sim1\text{--}2$ kpc, with the dashed gray line denoting 1~kpc for reference. However, the DFM deposits high-angular-momentum gas, which promotes star formation out to larger radii, and $r_{1/2}$ can grow much larger than $r_c$. 
Instead, the SIDM $r_c$ range from $\sim 1\text{--}4$~kpc, and the DFM allows stars to form at larger radii, which can be comparable to or larger than dark matter core sizes. The larger core sizes in the presence of disk formation shifts the relationship between $r_c$ and $r_{1/2}$ closer to $1:1$  in the VX-SIDM model. 

The combined impact of the two morphological classes results in a cosmology-dependent $r_c$--$r_{1/2}$ relationship at low-$z$. At $z = 0.3$, the slope derived from linear regression of the $r_c$--$r_{1/2}$ relationship in SI50 (1.19) sits between VX-SIDM (1.07) and CDM (1.41). This is because the SI50 model reaches maximum core formation and starts to collapse by $z = 0.3$, so core sizes decrease relative to VX-SIDM while still remaining larger than CDM. At $z = 0$, the $r_c$--$r_{1/2}$ slope is steeper compared to that at $z=0.3$ for both CDM and VX-SIDM (1.52 and 1.18, respectively).  For CDM, the core size remains essentially constant between these two redshifts, but $r_{1/2}$ becomes slightly larger for the disk systems. For VX-SIDM, disk sizes continue to grow faster than core sizes for the most extreme disk galaxies.

The state-of-the-art Massive SIDM-SIDM simulations enabled the first study of strong SIDM–baryon co-evolution in isolated, core-collapsing dwarf galaxies in a fully cosmological context. However, increasing simulation resolution may accelerate core collapse in SI50~\cite{2026arXiv260602566S}, which can produce smaller irregular galaxies during collapse and alter disk galaxy sizes depending on the DFM timing. Thus, these findings motivate targeted, high-resolution follow-up to more precisely identify specific core and half-light radii in the SI50 case.

In conclusion, this work used the Massive SIDM-SIDM zoom-ins, a novel suite of fully cosmological hydrodynamic simulations with varying SIDM strengths. These simulations demonstrate that galaxy morphology and size can be used to test the collisionality of dark matter. Dwarf galaxies with irregular morphologies show a nearly $1:1$ relation between dark matter and half-light radius, but with a distribution of sizes that depends on the dark matter model. The stellar size of disk dwarf galaxies correlates with dark matter core size in a model-dependent way. The Legacy Survey of Space and Time (LSST) survey is expected to detect millions of dwarf galaxies up to moderate redshifts~\cite{Hafezianzadeh2026}, enabling massive statistical studies of dwarf morphology evolution~\cite{Robertson2019Nat, Ivezi2019}. By demonstrating that the underlying dark matter microphysics directly shapes galaxy sizes and morphologies, our work provides a bridge between dark matter physics and observed galaxy properties that can be critically tested in the era of Rubin Observatory's LSST. 

\section{Simulations and Methods}\label{methods}
All simulations are run in a fully cosmological context and adopt the fiducial galaxy formation model used in the Massive Marvelous Dwarf suite presented in~\cite{Cruz2026}. The suites use the N-body$+$Smooth Particle Hydrodynamics~(SPH) code Charm Nbody Gravity solver~\citep[{\sc ChaNGa},][]{Menon2015}, which adopts and extends the hydrodynamic physics from {\sc Gasoline2} \citep{Wadsley2004, Wadsley2017} and has also been modified to use the {\sc charm}$++$ runtime code for load balancing and communication. Each zoom-in simulation is run from an initial condition~\cite{Pontzen2008} selected from the (25 Mpc)$^3$ box, Romulus25~\citep{Tremmel2017}. We select ``isolated" halos that are at least 1.5~Mpc away from any galaxy with $M_* > 10^{10}$~M$_{\odot}$ at redshift $z = 0.1$. Of the initial CDM Massive Dwarf simulations, we rerun 24 using the VX-SIDM model; of these, 9 are resimulated for the SI50 model as well. All simulations assume a Planck cosmology~\citep[$\Omega_{\Lambda} = 0.6914$, $\Omega_{\mathrm{m}} = 0.3086$, $\Omega_{\mathrm{b}} = 0.04825$,  $\sigma_8 = 0.8288$, $n_s = 0.9611$, $h = 0.6777$,][]{Plank2014}. All simulations use a spline force softening of $\epsilon = 87$~pc, an initial gas mass resolution of 3300~M$_{\odot}$, an initial stellar mass resolution of 990~M$_{\odot}$, and a dark matter mass resolution of $1.8 \times 10^4$~M$_{\odot}$.
\newline

\noindent {\bf Star formation, feedback, and black hole subgrid modeling} 

\noindent The same resolution and baryonic galaxy formation models are used in all three dark matter cosmologies. For a detailed description of the star formation, feedback, and black hole models, see~\cite{Cruz2026}. \newline

\noindent {\bf Self-interacting dark matter models}

\noindent 
We consider two models of self-interacting dark matter. In our first model, denoted VX-SIDM, we focus on the non-perturbative regime, where the interaction between particles at relative velocity $v_{\rm rel}$ is described in the classical limit ($m_\chi v_{\rm rel}/m_\phi \gg 1$)~\cite{Feng2010, Loeb2011, Tulin2013}, where $m_\chi$ is the dark matter mass and $m_\phi$ is the mediator mass. Following~\cite{Vogelsberger2012, Vogelsberger2014}, the transfer cross section $\sigma_T$ is parameterized with $\beta = 2\alpha_\chi \,m_{\phi}/(m_{\chi}\,v_{\rm rel}^2) \equiv \pi\,v_0^2/v_{\rm rel}^2$, where $\alpha_\chi$ is the dark matter--mediator coupling strength.
The VX-SIDM model assumes a maximum transfer cross section of $\sigma_{T}^{\rm max} \equiv 22.7/m_{\phi}^2 = 50$~cm$^2$/g 
and characteristic velocity of $v_{\rm 0} = 35$~km/s. The VX-SIDM model overlaps with the transfer cross section range motivated by the observed densities of Milky Way satellites in~\cite{Correa2021}. However, the VX-SIDM model holds in the classical limit rather than the Born limit ($\alpha_{\chi}m_{\chi}/m_{\phi} \ll 1$), as assumed in~\cite{Correa2021}. Further, the model derived in~\cite{Correa2021} is subject to a number of uncertainties, including the impact of ram-pressure stripping, and uncertainties in the mass of the Milky Way, orbital velocities of its satellites, and radial heat conduction in the long-mean-free-path regime of the interacting system.

The SI50 suite adopts a constant transfer cross section of $\sigma_T = 50$~cm$^2$/g. Utilizing the scaling relationship of~\cite{Zeng2022}, we selected $\sigma_T$ such that a fraction of the simulated systems would core collapse within a Hubble time if they were dark-matter-only. However, baryon mass speeds up collapse~\cite{Zhong2023} while feedback slows it down~\cite{Kong2026}. Our SI50 systems tend to reach and surpass their maximal core phase within a Hubble time and more quickly with baryons than in dark-matter-only, indicating that the baryonic mass speed-up dominates over feedback in our model. To calculate $\sigma_T$, we determine the concentration of each CDM halo using the method outlined in Appendix C of~\cite{Cruz2026}.  In particular, we determine the scale radius $r_{-2}$, the location where the log-density slope $\frac{{\rm dlog}~\rho(r)}{{\rm dlog}~r}  = - 2$, and then set $c_{200, c} = r_{\rm 200,c}/r_{-2}$, where $r_{200,c}$ is the virial radius with an overdensity of 200 times the critical density of the Universe. To agree with observational constraints, the physical SIDM cross section should be velocity-dependent~\cite{Tulin2013, Correa2021}; thus, the constant SI50 model is a simplifying assumption that mimics an effective cross section of $50$~cm$^2$/g within the simulated mass range. Motivated by kinetic theory, the effective cross section maps a velocity-dependent cross section onto a constant cross section by integrating out the velocity and angular dependence of a full cross section model~\cite{Yang2022}.
\newline

\noindent {\bf Halo finding, centering, tracing, and merging}

\noindent Halos are identified using the AMIGA halo finder \citep[AHF;][]{Knollmann2009}. The virial radius, $r_{\rm vir}$, of a halo is the radius at which the density corresponds to the redshift-dependent virial overdensity derived from the spherical collapse model~\cite{Bryan1998} times the critical density of the Universe. Halos are centered using the shrinking-sphere procedure~\citep{Power2003}. Galaxies are rotated into the x--y plane to their ``face-on" orientation such that the angular momentum vector of the gas within 5~kpc of the halo center points in the z-direction. While this method is used independent of morphology, for the disk galaxies specifically, the z-component of the total angular momentum is well aligned with the z-direction \citep{Geda2026}. Halo merger tree construction and sub-halo  tracing through time utilizes the ``voter and graveyard'' methods of \citep{Ivkovic2022, GedaTrees2025}. In this work, mergers are defined as secondary galaxies that enter the host's virial radius with a mass ratio of at least $1:30$ and lose at least 90$\%$ of their in-fall dark matter mass to the host. Following~\cite{Geda2026}, the DFM is the merger with the largest orbital angular momentum and $t_{\scriptscriptstyle{\rm DFM}}$ is the time when the merger's angular momentum is maximal. All other merger times correspond to when the merging satellite first enters the virial radius of the host. 
\newline

\noindent {\bf Stellar sizes}

\noindent We define the stellar size of a galaxy to be the half-light radius of its stars, $r_{1/2}$, in the Johnson $K$-band using cylindrical coordinates. All stellar sizes are measured in the ``face-on" orientation. The $K$-band is chosen to mimic SPARC~\citep{Lelli2016}. {\sc pynbody} computes luminosities of star particles in the $K$-band using the Padova simple stellar populations~\citep{Marigo2008}, constructs a galaxy's total luminosity profile, and identifies the galaco-centric radius that encloses half of that luminosity. In~\cite{Cruz2026}, we verified the stellar size measurements in CDM at $z = 0$ by generating mock $K$-band images of the CDM galaxies using {\sc pynbody} and measured their Petrosian profiles~\cite{Petrosian1976} with {\sc PetroFit}~\cite{Geda2022}. Measurements using SDSS parameters~\cite{Strauss2002} yielded nearly identical values to the Petrosian half-light radii and the {\sc pynbody}-derived $r_{1/2}$. See~\cite{Cruz2026} for additional size tests at $z = 0$ in the CDM Massive Dwarfs. 
\newline

\noindent 

\noindent {\bf Density profiles and core sizes}

\noindent We calculate spherically averaged dark matter density profiles between the convergence radius $r_{\rm conv}$ and $r_{\rm vir}$ using {\sc pynbody}~\cite{pynbody} in all three dark matter cosmologies. Below the convergence radius, $r_{\rm conv}$, two-body scattering can alter mass distributions~\citep{Power2003} in CDM. The convergence radius interior to which artificial heating is expected is approximated as $r_{\rm conv} = 0.055\times l$,  where $l = L_{\rm box}/N_{\rm part}^{1/3}$ is the mean interparticle spacing in physical units for a simulation with box length $L_{\rm box}$ and $N_{\rm part}$ total particles~\cite{LSB2019}. In the Massive Dwarfs, $r_{\rm conv}\sim0.45~{\rm kpc}$. Because of strong dark matter self-scattering at low-$z$ in the SI50 model, the rate of interactions may exceed that of artificial two-body scattering, and thus the convergence radius may be lower than the typical convergence radius in CDM. For uniformity, the CDM convergence radius is used in all three dark matter cosmologies, but may be an upper limit in the SI50 case in the collapse phase. The dark matter core size, $r_c$, is defined as the radius where the maximum resolved density at $r_{\rm conv}$ falls by a factor of two. The convergence radius effectively puts a lower limit of 0.45~kpc on the attainable dark matter core size. 

\noindent



\appendix

\setcounter{equation}{0}
\setcounter{figure}{0} 
\setcounter{table}{0}
\renewcommand{\theequation}{A\arabic{equation}}
\renewcommand{\thefigure}{A\arabic{figure}}
\renewcommand{\thetable}{A\arabic{table}}
\renewcommand{\thesection}{A\arabic{section}}

\section*{Appendix A: Co-evolution figures for all simulated galaxies} \label{sec:coevol_A}

\begin{figure}[H]
\centering
\includegraphics[width=1\linewidth]{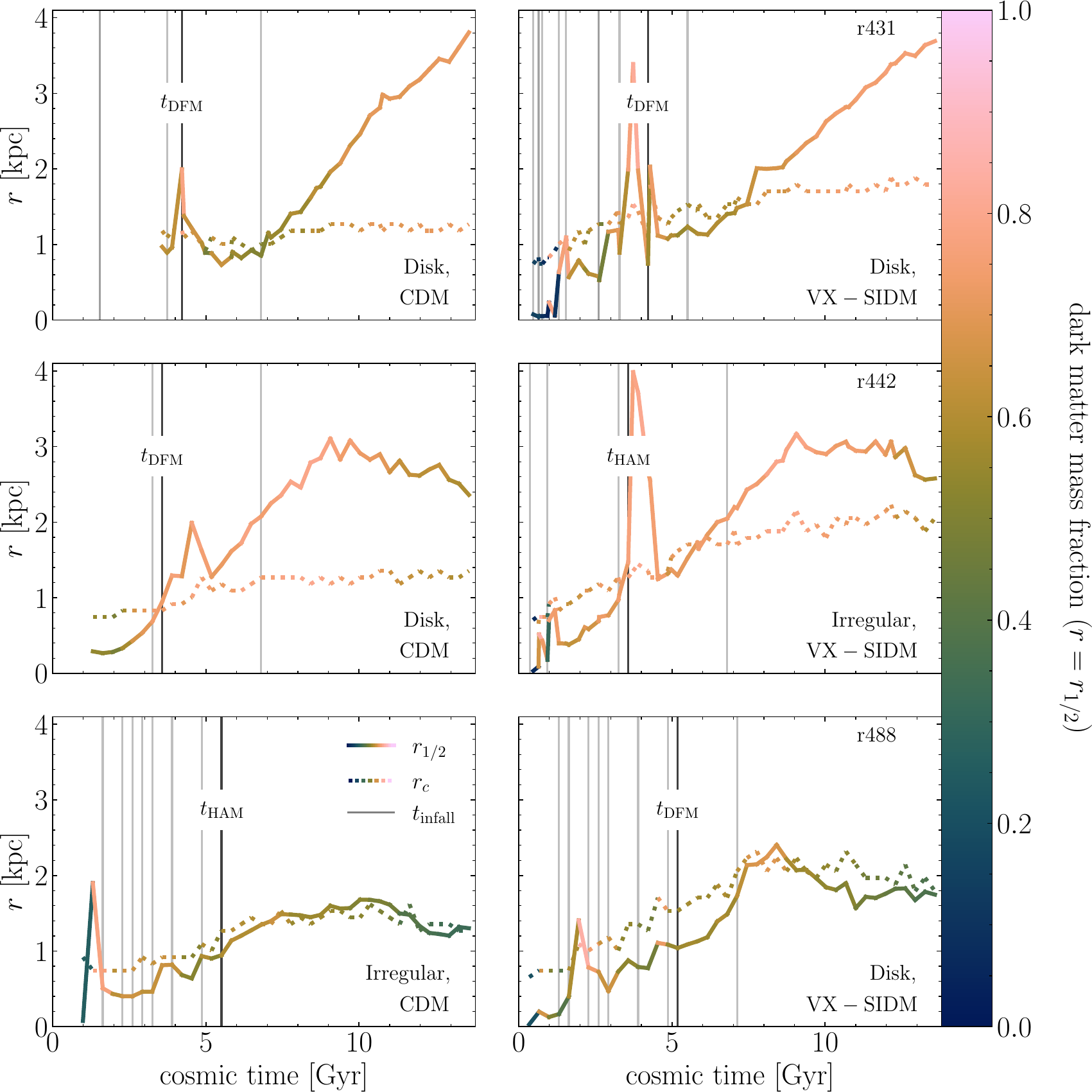}
\caption{The time evolution of dark matter core size ($r_c$, dotted lines) and galaxy half-light radius ($r_{1/2}$, solid lines), with tracks colored by the dark matter mass fraction within $r_{1/2}$. Results are shown for r431~(top row), r442~(middle row) and r488~(bottom row). In CDM, r431 is not traced to earlier times because simulation outputs were lost. Each cosmology is represented in a given column: CDM~(left) and VX-SIDM~(right). Vertical black lines denote the time of disk-forming merger, $t_{\rm DFM}$, and highest-angular-momentum merger, $t_{\rm HAM}$, for the disk and irregulars, respectively (see Methods Section~\ref{methods}). Vertical gray lines indicate the infall times ($t_{\rm infall}$) for all other mergers. In r431 and r442, dark matter dominates or exists in roughly the same proportion as baryons within $r_{1/2}$, except at early times. Instead, baryons dominate at late times in r488. In all cases, the minimum obtainable core size is limited to the CDM $r_{\rm conv} = 0.45$~kpc.
} \label{fig:size_evol1}
\end{figure}

\begin{figure}[H]
\centering
\includegraphics[width=1\linewidth]{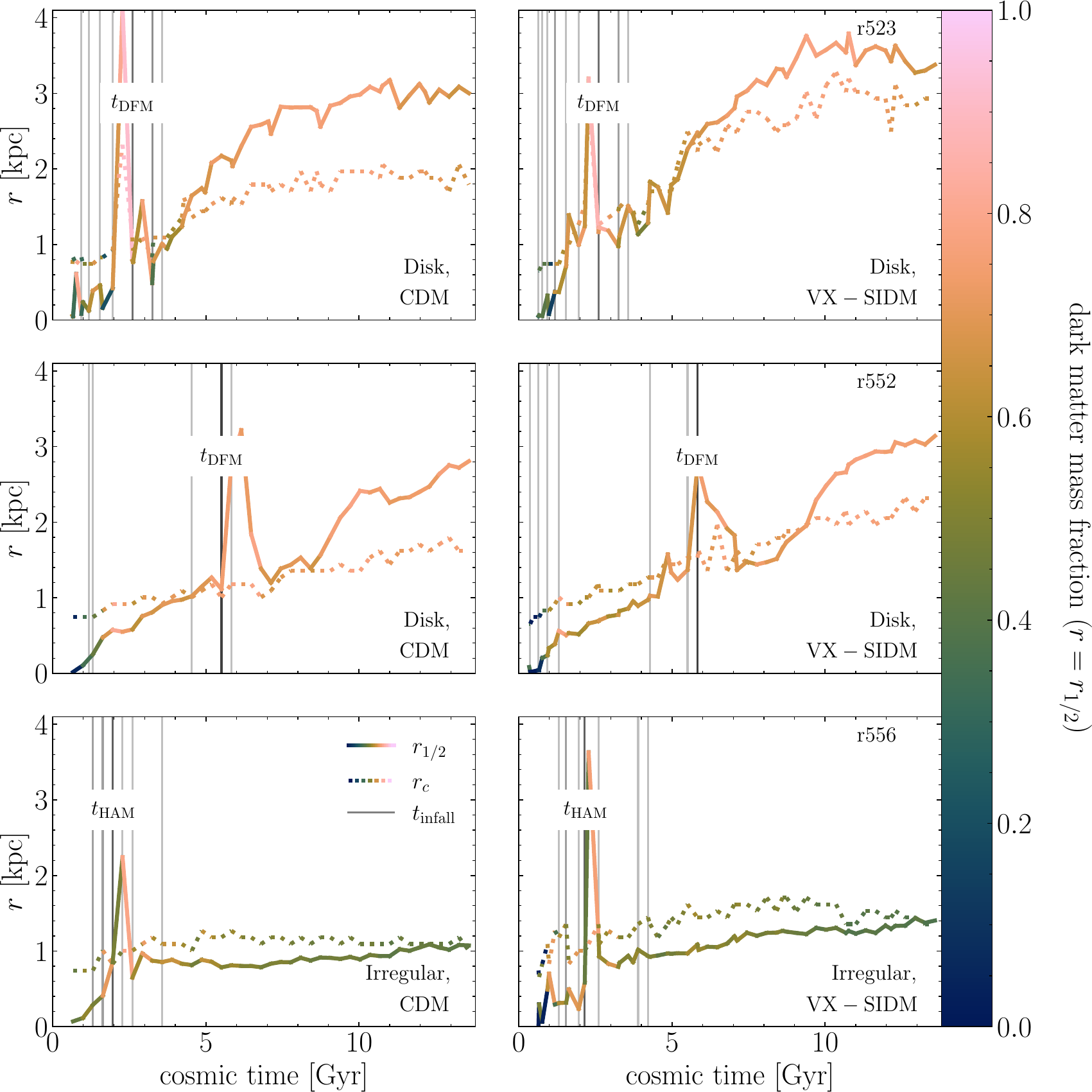}

\caption{Same as Figure~\ref{fig:size_evol1}, except for r523~(top row), r552~(middle row) and r556~(bottom row). In r523 and r552, dark matter dominates or exists in roughly the same proportion as baryons within $r_{1/2}$, except at early times. In contrast, baryons dominate at late times in r556, which is one of the most compact galaxies in the sample. } \label{fig:size_evol2}
\end{figure}

\begin{figure}[H]
\centering
\includegraphics[width=1\linewidth]{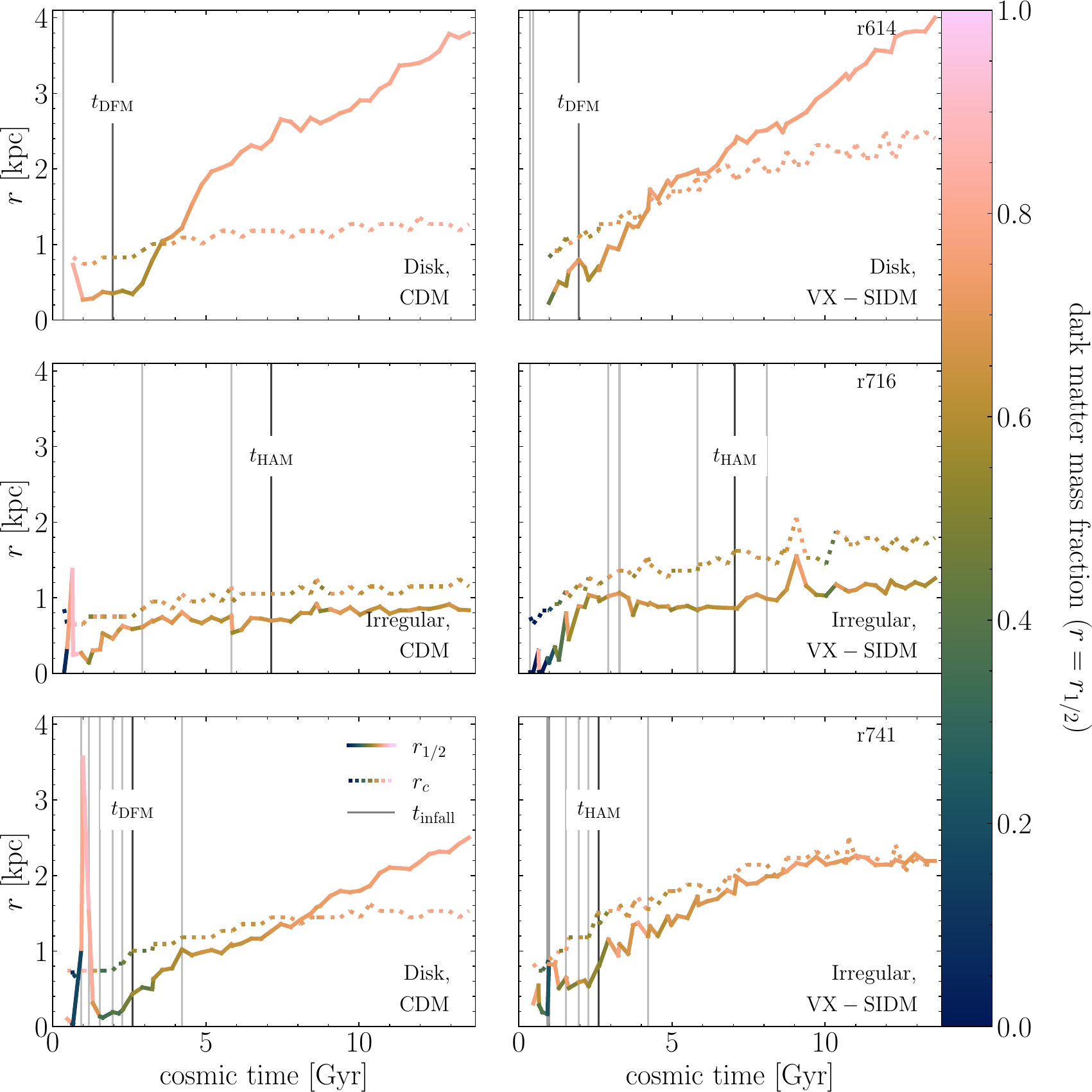}
\caption{Same as Figure~\ref{fig:size_evol1}, except for r614~(top row), r716~(middle row) and r741~(bottom row). In all cases, dark matter dominates or exists in roughly the same proportion as baryons within $r_{1/2}$, except at early times.} \label{fig:size_evol3}
\end{figure}

\begin{figure}[H]
\centering
\includegraphics[width=1\linewidth]{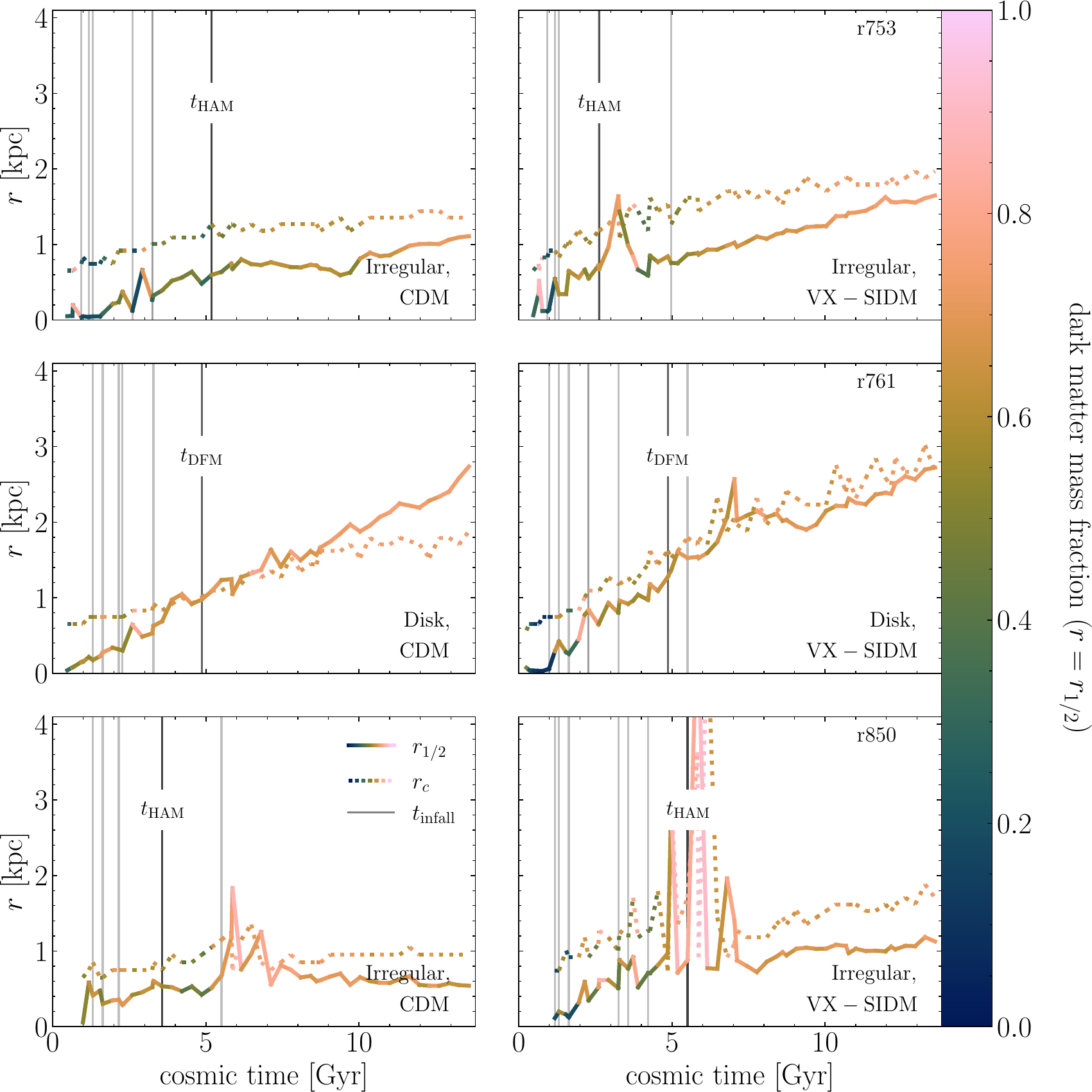}
\caption{Same as Figure~\ref{fig:size_evol1}, except for r753~(top row), r761~(middle row) and r850~(bottom row).
 In all cases, dark matter dominates or exists in roughly the same proportion as baryons within $r_{1/2}$, except at early times.  } \label{fig:size_evol4}
\end{figure}

\begin{figure}[H]
\centering
\includegraphics[width=1\linewidth]{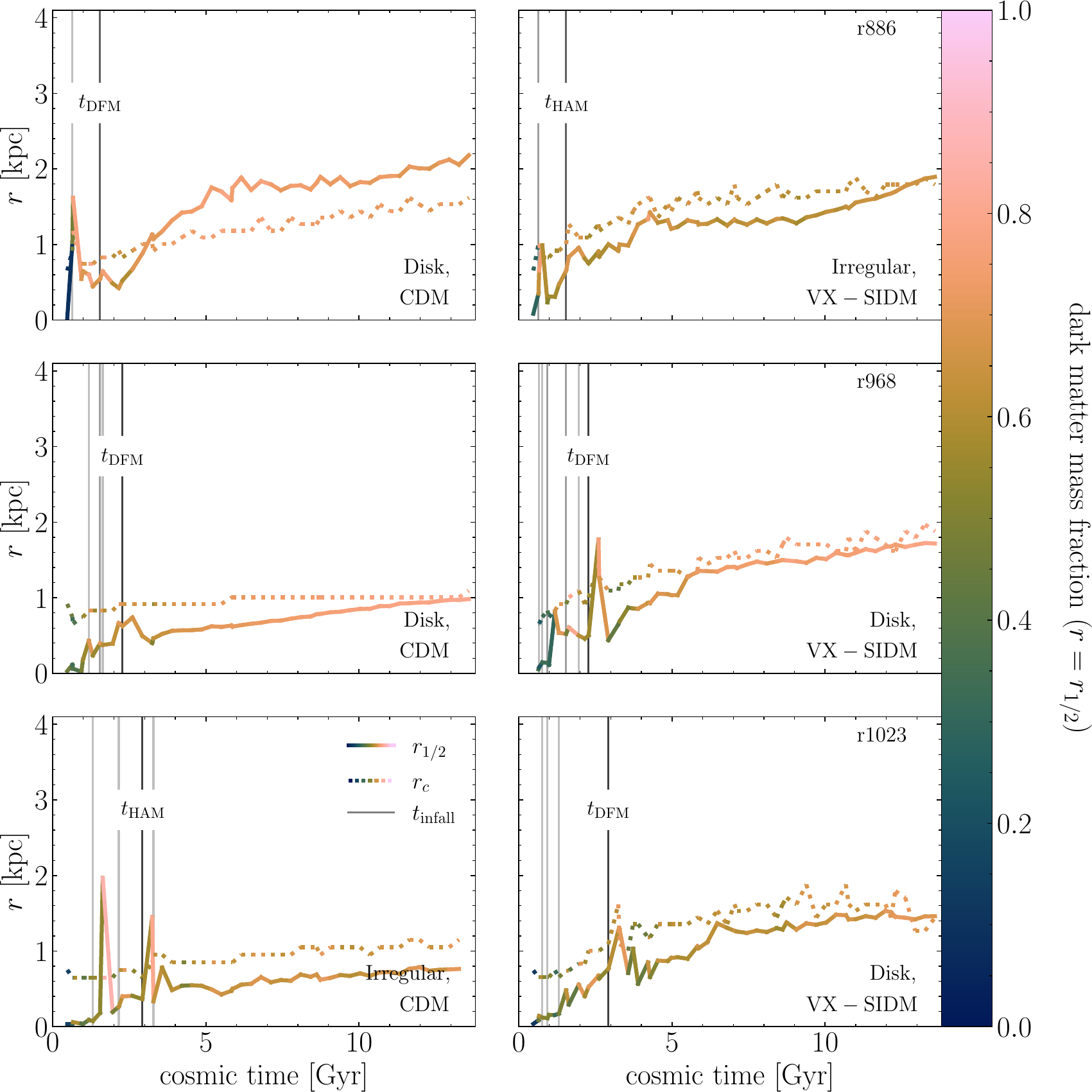}
\caption{Same as Figure~\ref{fig:size_evol1}, except for r886~(top row), r968~(middle row) and r1023~(bottom row).
 In all cases, dark matter dominates or exists in roughly the same proportion as baryons within $r_{1/2}$, except at early times.} \label{fig:size_evol5}
\end{figure}

\begin{figure}[H]
\centering
\includegraphics[width=1\linewidth]{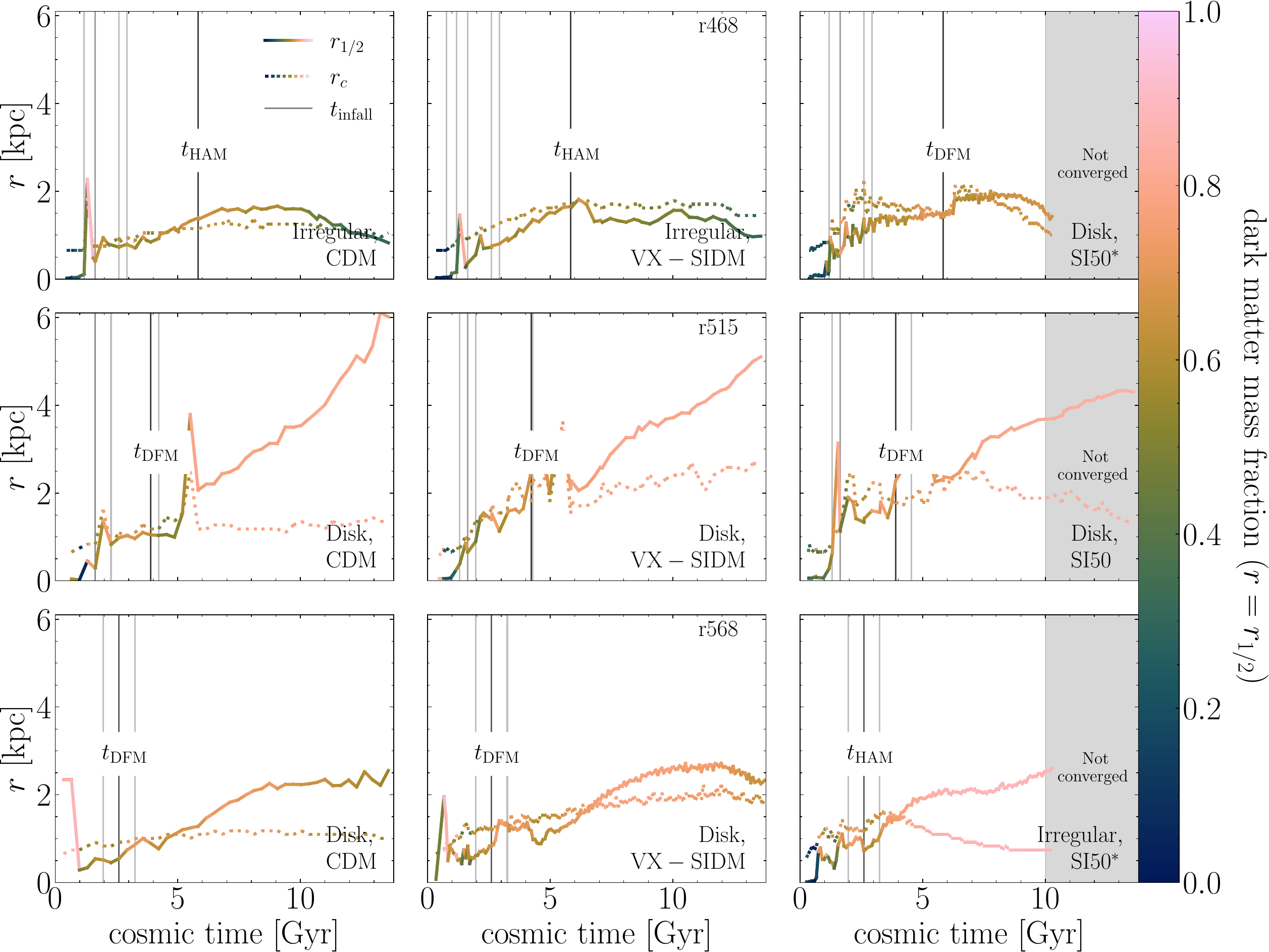}
\caption{Same as Figure~\ref{fig:size_evol1}, except for r468~(top row), r515~(middle row) and r568~(bottom row).  Additionally, results for the SI50 cosmology are provided in the third column. In the SI50 model, r468 and r568 use the CDM mergers, this is specified with a star on the cosmology label. The gray shaded region represents the time beyond which the central density evolution is no longer converged for SI50.  In r515 and r568, dark matter dominates or exists in roughly the same proportion as baryons within $r_{1/2}$, except at early times. In contrast, baryons dominate at late times in r468, which is one of the most compact galaxies in the sample. $r468$ was not run passed $z = 0.3$ because core-collapse is not converged past that point.} \label{fig:size_evol6}
\end{figure}

\begin{figure}[H]
\centering
\includegraphics[width=1\linewidth]{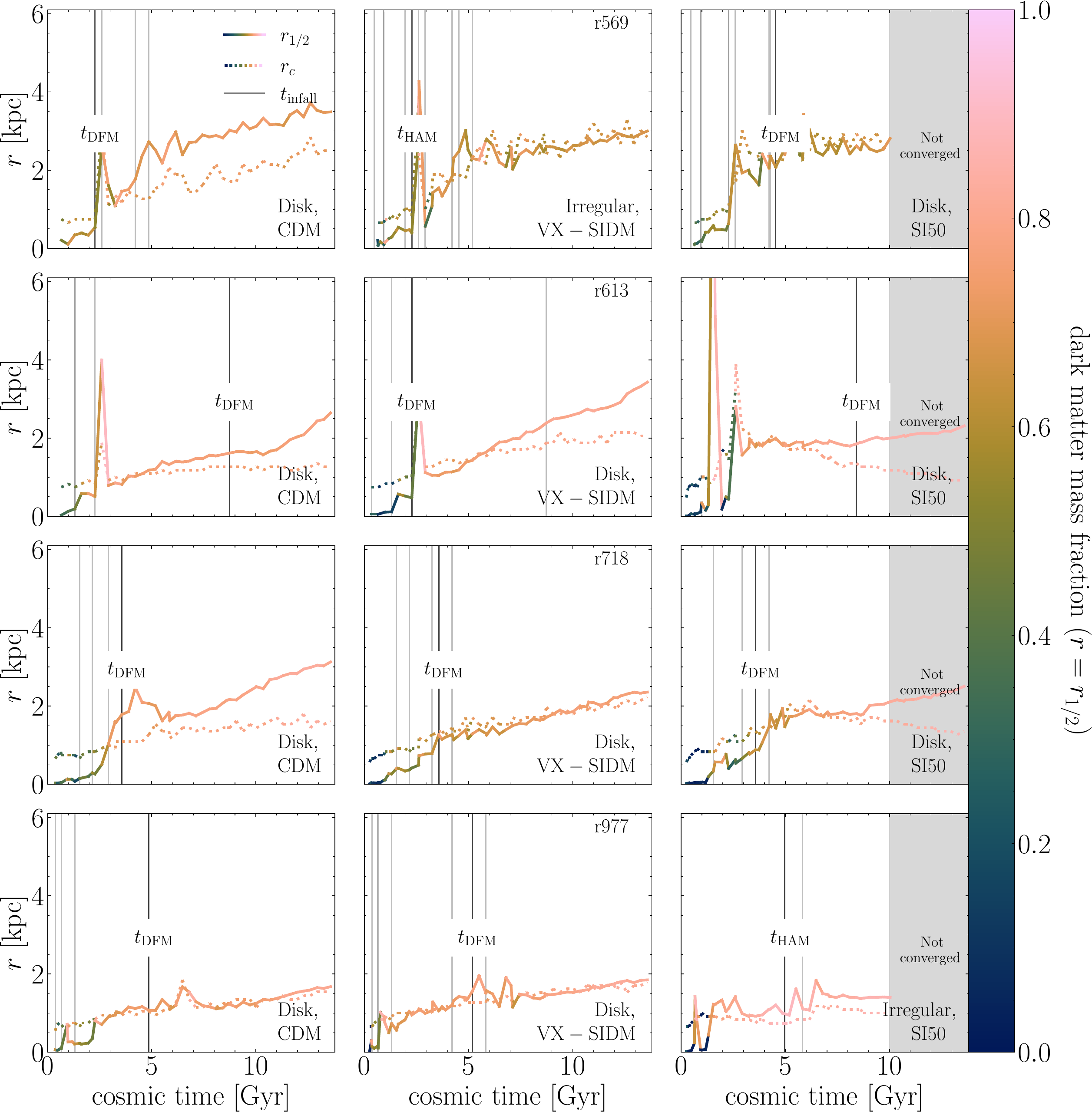}
\caption{Same as Figure~\ref{fig:size_evol6}, except for r569~(top row), r613~(second row), r718~(third row), r977~(bottom row). In all cases, dark matter dominates or exists in roughly the same proportion as baryons within $r_{1/2}$, except at early times.} \label{fig:size_evol7}
\end{figure}

The main text uses a representative irregular~(r618) and disk~(r634) galaxy from our simulation suite to demonstrate the co-evolution between $r_c$ and $r_{1/2}$ and how it is affected by disk growth. Figures~\ref{fig:size_evol1}--\ref{fig:size_evol7} present the $r_c$ and $r_{1/2}$ co-evolution plots for all the simulated galaxies used in this study. We provide the results across the CDM and VX-SIDM models, as well as the SI50 model when those galaxies exist. Each row in Figures~\ref{fig:size_evol1}--\ref{fig:size_evol7} provides the results for a single galaxy, whose name is indicated in the top-right of the VX-SIDM panel.  Each panel is formatted similar to Figure~2 in the main text. The vertical black line indicates the time for the disk-forming merger, $t_{\rm DFM}$, for the disk systems, or the highest-angular-momentum (HAM) merger, $t_{\rm HAM}$, for the irregulars. All other mergers are indicated by the vertical gray lines at the time when they first enter the host galaxy's virial radius ($t_{\rm infall}$, infall time). Instead, the HAM and DFM are taken when the orbital angular-momentum of the satellite is maximized. Some of the SI50 satellites have not fully merged (e.g., they have not lost 90$\%$ of their infall dark matter to the host) by $z = 0.3$ based on how mergers are defined (see Methods Section~\ref{methods}). In these cases, the CDM mergers are used as a proxy. Previous studies have shown that SIDM can speed up merger times by $\sim 10\%$~\cite{Kim2017}, however simulated SI50 systems run to $z = 0$ (r515, r613, and r718) show similar merger histories between CDM and SI50. The systems where the CDM mergers are taken for the SI50 model are r468, and r568, r634; this is denoted in Figure~\ref{fig:size_evol6} with stars on the cosmology as SI50$^*$.

In general, all the galaxies follow the same behavior as the two representative systems in the main body: the dark matter and stars co-evolve through time for the irregular galaxies.  For disk systems, this co-evolution persists until the DFM, after which $r_{1/2}$ and $r_c$ decouple.

Note that the original morphology in CDM is not always preserved in VX-SIDM. Out of the 24 simulated galaxies, 3 that form disks in CDM are irregular in VX-SIDM. On the other hand, 2 galaxies that are irregular in CDM form disks in VX-SIDM. This may be due in part to numerical chaos~\cite{Keller2019, Genel2019}, and the definition chosen for the disk classification. A table of the median, normalized, specific angular momentum in the z-direction ($j_{s, z}/j_s$) of the stellar population for each galaxy is presented in Extended Data Table \ref{tab:jzj}.

\begin{table}[ht]
  \centering
  \caption{For each simulation (first column) the morphology type using CDM at $z=0$ (second) and median specific angular momentum in the z-direction in CDM at $z=0$ (third), CDM at $z=0.3$ (fourth), VX-SIDM at $z=0$ (fifth), VX-SIDM at $z=0.3$ (sixth) and SI50 at $z=0.3$ (seventh) are tabulated. Superscripts denote morphology changes between cosmologies based on $j_{s, z}/j_z$ definition, otherwise the original morphology remains.}
  \label{tab:jzj}
  \begin{tabular}{ c c c c c c c }
    \hline
    Simulation & Morphology & ${\rm CDM}$ & ${\rm CDM}$ & ${\rm VX-SIDM}$ & ${\rm VX-SIDM}$  & ${\rm SI50}$ \\
    &  & $j_{s,z}/j_s$ & $j_{s,z}/j_s$  & $j_{s,z}/j_s$ & $j_{s,z}/j_s$ & $j_{s,z}/j_s$ \\
     &  & $z=0$ & $z=0.3$  & $z=0$ & $z=0.3$ & $z=0.3$ \\
    \hline
    r431 & disk & 0.94 & 0.85 & 0.93 & 0.83 & -- \\
    \hline
    r442 & disk & 0.9 & 0.87 & 0.9 & 0.83 & -- \\
    \hline
    r468 & irregular\textsuperscript{\it a} & 0.21 & 0.77 & 0.17 & 0.47 & 0.79 \\
    \hline
    r488 & irregular\textsuperscript{\it b} & -0.03 & 0.03 & 0.59 & 0.04 & -- \\
    \hline
    r515 & disk & 0.93 & 0.85 & 0.91 & 0.85 & 0.88 \\
    \hline
    r523 & disk & 0.9 & 0.85 & 0.88 & 0.87 & -- \\
    \hline
    r552 & disk & 0.59 & 0.51 & 0.62 & 0.69 & -- \\
    \hline
    r556 & irregular & 0.1 & 0.0 & 0.12 & 0.03 & -- \\
    \hline
    r569 & disk\textsuperscript{\it c} & 0.84 & 0.69 & 0.46 & 0.53 & 0.7 \\
    \hline
    r613 & disk & 0.81 & 0.72 & 0.82 & 0.76 & 0.85 \\
    \hline
    r614 & disk & 0.91 & 0.9 & 0.87 & 0.87 & -- \\
    \hline
    r618 & irregular & 0.07 & -0.01 & 0.07 & 0.17 & 0.51 \\
    \hline
    r634 & disk\textsuperscript{\it d} & 0.67 & 0.23 & 0.55 & 0.46 & 0.92 \\
    \hline
    r716 & irregular & -0.02 & 0.07 & 0.05 & -0.01 & -- \\
    \hline
    r718 & disk & 0.85 & 0.83 & 0.63 & 0.71 & 0.81 \\
    \hline
    r741 & disk\textsuperscript{\it c} & 0.87 & 0.84 & 0.38 & 0.64 & -- \\
    \hline
    r753 & irregular & 0.27 & 0.05 & 0.5 & 0.4 & -- \\
    \hline
    r761 & disk & 0.73 & 0.76 & 0.81 & 0.77 & -- \\
    \hline
    r850 & irregular & 0.02 & 0.01 & -0.04 & 0.04 & -- \\
    \hline
    r886 & disk\textsuperscript{\it c} & 0.86 & 0.81 & 0.38 & 0.2 & -- \\
    \hline
    r968 & disk & 0.61 & 0.53 & 0.67 & 0.75 & -- \\
    \hline
    r977 & disk & 0.79 & 0.7 & 0.84 & 0.8 & 0.91 \\
    \hline
    r1023 & irregular\textsuperscript{\it b} & 0.04 & 0.08 & 0.75 & 0.74 & -- \\
    \hline
  \end{tabular}
  \footnotetext[a]{Irregular in VX-SIDM, disk in SI50 and CDM at $z = 0.3$}
  \footnotetext[b]{Irregular in CDM, disk in VX-SIDM at $z = 0$}
  \footnotetext[c]{Disk in CDM, irregular in VX-SIDM at $z = 0$}
  \footnotetext[d]{Irregular in CDM and VX-SIDM, disk in SI50 at $z = 0.3$}
\end{table}

\setcounter{equation}{0}
\setcounter{figure}{0} 
\setcounter{table}{0}
\renewcommand{\theequation}{B\arabic{equation}}
\renewcommand{\thefigure}{B\arabic{figure}}
\renewcommand{\thetable}{B\arabic{table}}
\renewcommand{\thesection}{B\arabic{section}}

\section*{Appendix B: SIDM convergence tests} \label{sec:convergeB}

ChaNGa uses the SIDM implementation introduced in~\citet{Fry2015}, which is largely based on~\cite{Vogelsberger2012}. The original code has been updated to include the following dynamic time-stepping scheme: 

The acceleration timestep for the $i^{\rm th}$ SIDM particle is defined as ${\rm d}t_{a,i} = \eta \sqrt{\epsilon/|a_i|}$~\cite{Power2003}, where $\epsilon$ is the gravitational softening, $a_i$ is the acceleration, and $\eta$ is a constant (set to 0.185 for CDM). The SIDM interaction probability timestep is ${\rm d}t_{{\rm SIDM},i}= \kappa (\sigma_{T} \, \rho_{{\rm DM}, i} \, v_{{\rm rel}, i})^{-1}$,  where $\sigma_T$ is the SIDM transfer cross section, $\rho_{{\rm DM}, i}$ is the cubic-spline kernel-smoothed dark matter density of the $k=64$ nearest neighbors, $v_{{\rm rel}, i}$ is the local velocity dispersion, and $\kappa$ is a free parameter used to set the limiting interaction probability. For each SIDM particle, the chosen timestep is the minimum between ${\rm d}t_{a,i}$ and ${\rm d}t_{{\rm SIDM}, i}$.

\begin{figure}[H]
\centering
\includegraphics[width=0.75\linewidth]{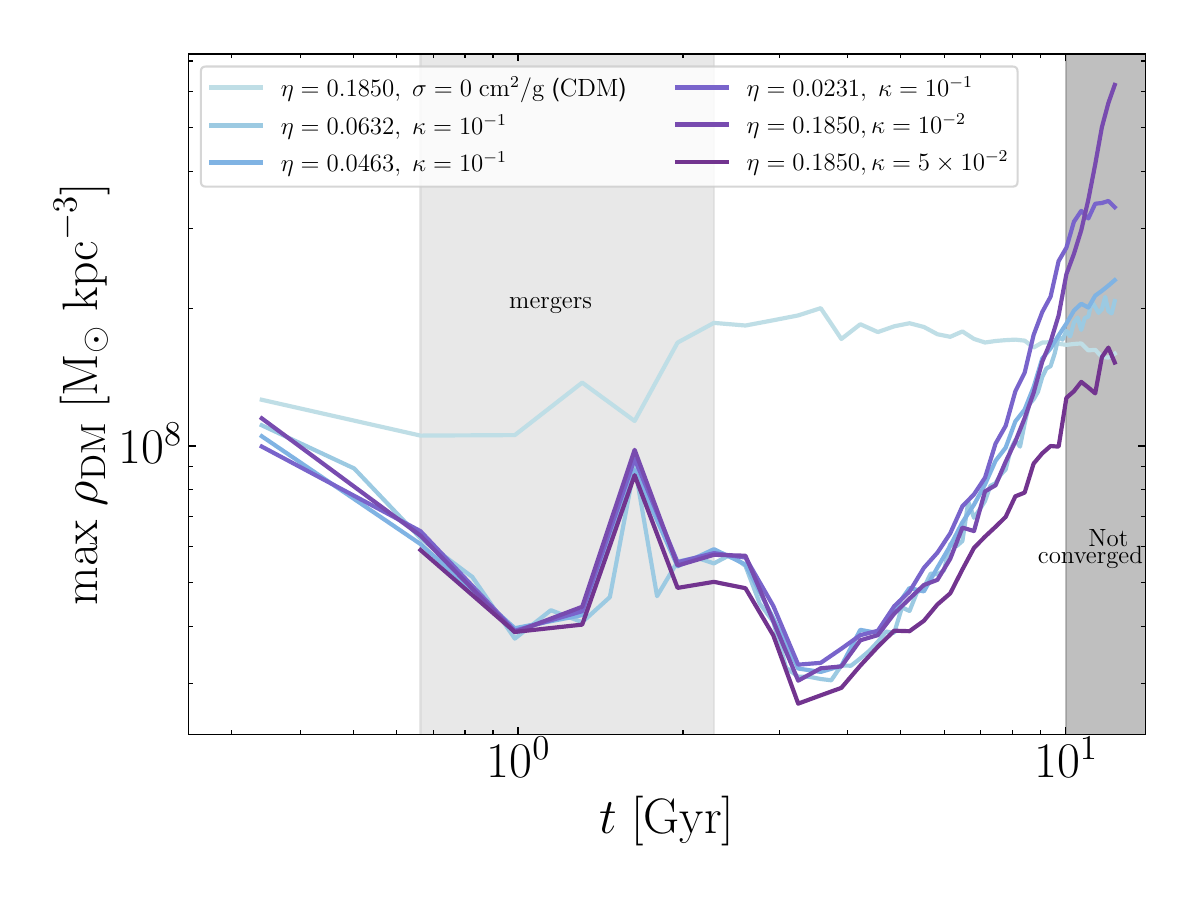}
\caption{Central dark matter density versus time for varied time stepping parameters $\kappa$ and $\eta$. The fiducial values used for this study are $\kappa = 0.01$ and $\eta = 0.185$. The light gray shaded region represents the time when mergers are influencing the central density. The dark-gray shaded region marks where deviations in central density evolution persist at late times, near the converging curves surrounding the fiducial run.} \label{fig:calib}
\end{figure}

We re-run the initial condition with the smallest expected collapse cross section (r568) based on the core-collapse time-scaling relationship from dark-matter-only simulations (see Methods Section~\ref{methods}), varying the values of $\eta$ and $\kappa$. We vary $\kappa$ between $10^{-2}$, $5 \times 10^{-2}$ at $\eta = 0.185$ and $\eta$ between $0.0231$, $0.0463$, $0.0632$ at $\kappa = 10^{-1}$.  The results are summarized in Figure~\ref{fig:calib}, which demonstrates the evolution of the central density within 1~kpc as a function of time. While the simulated halos continue to grow in density beyond $z=0.3$, the rate of change in the central density varies depending on the time-stepping scheme, as has been previously found in both isolated and cosmological simulations~\cite{Mace2024, Palubski2024, Fischer2024, 2026arXiv260602566S}. We select $\kappa = 10^{-2}$ and $\eta = 0.185$ as fiducial values for this study. This choice balances convergence considerations, with computational requirements: decreasing $\kappa$ by a factor of 2 increases the computation cost by a factor of $\sim~5$. The fiducial values of $\kappa$ and $\eta$ allow the most extreme system in our suite to converge at late times, minimizing secondary scatters in this one system, as well as in the others considered. We do not attempt to provide cross section constraints, and instead we show expected trends depending on whether an object is left in the core or collapse phase at low-$z$. To exact precision cross-section constraints, additional convergence tests and feedback model testing is required. 

\backmatter
\bmhead{Acknowledgements}

The authors thank Manoj Kaplinghat, Stacy Kim, and Carton Zeng for helpful discussions on this work. AMB acknowledges support from National Science Foundation (NSF) grant AST-2306340 and from grant FI-CCA-Research-00011826 from the Simons Foundation. ML is supported by the Department of Energy under Award Number DE-SC0007968, the Simons Investigator in Physics Award, and grant SFI-FI-CCA-RESEARCH-00019731 from the Simons Foundation. AHGP is supported by NSF Grant No. AST-2008110 and AST-2510899. BWK acknowledges support provided by NASA through a grant from the Space Telescope Science Institute, through grant HST AR-17547. This work used Stampede2 at the Texas Advanced Computing Center through allocation MCA94P018 from
the Advanced Cyberinfrastructure Coordination Ecosystem: Services $\&$ Support program, which is supported by U.S. National Science Foundation grants 2138259, 2138286, 2138307, 2137603, and 2138296. Resources supporting this work were provided by the NASA High-End Computing Program through the NASA Advanced Supercomputing Division at Ames Research Center. Some of the simulations were performed using resources made available by the Flatiron Institute. The Flatiron Institute is a division of the Simons Foundation.

\bmhead{Author contributions}
All authors contributed substantially to this paper. Every author read and provided thorough comments on the manuscript. AC conceived the original idea for this work and led the analysis. RG contributed to the analysis of simulation data, including code development. AC and AMB collaborated to generate the ICs and run the CDM simulations. AC ran the SIDM simulations. ML, AMB, and AHGP supervised the project, helping to structure of the manuscript and  interpret the results. AC and TRQ developed the SIDM time-stepping code. BK developed the superbubble feedback model used. TRQ is a main builder of ChaNGa. 

\bmhead{Code availability}
The ChaNGA code was used to run the hydrodynamical simulations used in this work; the public version is available at \url{https://github.com/N-bodyShop/changa}. Pynbody and Tangos were used to obtain stellar and dark matter core sizes and to produce the mock UVI images and are publicly available at \url{https://github.com/pynbody/pynbody} and \url{https://github.com/pynbody/tangos}.  AHF was used to identify halos and is available at \url{http://popia.ft.uam.es/AHF/index.html}. 

\bmhead{Data availability}
The simulations are available upon reasonable request on a case-by-case basis.

\bibliography{main-4-2}

\end{document}